\documentclass[12pt]{article}
\pdfoutput=1
\usepackage{graphicx}
\usepackage{epstopdf}
\usepackage{amsmath}
\usepackage{amsfonts}
\usepackage{amssymb}
\usepackage{color}
\usepackage{mathrsfs}

\usepackage{graphicx} \usepackage{bm} \usepackage{epsfig}
\usepackage{amsmath, amssymb}
\usepackage{hyperref}
\usepackage{setspace}
\usepackage{color}
\usepackage{colordvi}
\usepackage{comment}
\usepackage{graphicx}

\begin{document}

\begin{center}

{\Large \bf Analytic approximations, perturbation theory,\\[0.3cm] effective field theory methods and their applications}\\[0.7cm]

{\large Vitor Cardoso${}^{1,2}$ and Rafael A. Porto${}^{3,4}$}
\\[0.7cm]

{\normalsize {\sl $^{1}$ CENTRA, Departamento de Fisica, Instituto Superior T\'ecnico,
Universidade de Lisboa, Avenida Rovisco Pais 1, 1049 Lisboa, Portugal.}}\\ 
\vspace{.3cm}

{\normalsize {\sl $^{2}$ Perimeter Institute for Theoretical Physics, Waterloo, Ontario N2L 2Y5, Canada.}}\\
\vspace{.3cm}

{\normalsize { \sl $^{3}$ School of Natural Sciences, Institute for Advanced Study, \\Olden Lane, 
Princeton, NJ 08540, USA.}}\\
\vspace{.3cm}

{\normalsize { \sl $^{4}$ Deutsches Elektronen-Synchrotron DESY, Theory Group, D-22603 Hamburg, Germany.}}\\
\vspace{.3cm}

\end{center}

\vspace{.8cm}

\hrule \vspace{0.3cm}
{\small  \noindent \textbf{Abstract} \\[0.3cm]
We summarize the parallel session B4: `Analytic approximations, perturbation theory effective field theory methods and their applications' and the joint session B2/B4: `Approximate solutions to Einstein equations: Methods and Applications', of the GR20 \& Amaldi10 conference in Warsaw, July 2013. The contributed talks reported significant advances on various areas of research in gravity.\\[0.3cm]
\noindent  \vspace{0.3cm}
\hrule

\newpage
\section{Summary}

The GR20 \& Amaldi10 meetings in Warsaw are the last ones of the series before the centennial of Einstein's theory of gravitation. After nearly one hundred years since the theory of General Relativity was discovered, Kerr's metric describing vacuum (neutral) rotating bodies, are among the few known exact solutions in four-dimensional, asymptotically flat space-times. (However, see \cite{kramer} for a comprehensive review.) The lack of (generic) solutions to the $N$-body problem in General Relativity highlights the importance of numerical techniques, perturbative methods such as the Post-Newtonian (PN) approximation for comparable masses or black hole perturbation theory for extreme-mass-ratio inspirals (EMRIs), as invaluable tools to solve for gravitational dynamics. 

These techniques are of paramount importance in light of the programme to directly observe Gravitational Waves (GWs) on Earth- and space-based detectors, which includes the computation of GW observables for compact binary systems, such as the phase and amplitude of the GWs produced as the system inspirals towards merger, at the moment still awaiting detection  by a network of laser interferometers including (advanced) LIGO/Virgo, and the next generation of GW observatories. The payoff of GW science will rely upon the use of accurate signal models ({\it aka} templates) from the most promising sources, and therefore analytic and/or combined numerical/analytic efforts for the study of binary dynamics are essential to extract the most information from the data, like measurements of masses and spins to high precision. The inspiral and merger of compact objects are also a natural laboratory to test gravity in the strong-field regime to an unprecedented level. Although not expected to prevent GW detection, the lack of sufficiently accurate templates and/or putative modifications of gravity in this regime may hinder parameter estimation and the ability to correctly map the contents throughout the universe, if not properly modeled. 

Finally, with all the above as motivation, the combination of perturbation techniques with numerical methods is providing novel insight into the structure of the non-linear field equations in the strong-field regime, including cosmic censorship violations and interesting connections with high-energy and particle physics.

\subsection{Dynamics in General Relativity:\\ Post-Newtonian, Self-force, EFT \& EOB methods}

For comparable mass non-rotating compact bodies the dynamics of the binary and GW phase have been derived up to 3.5PN order \cite{35pn}. This includes the computation of the binding energy (and equations of motion) of the orbit, and radiative multipole moments (including tail effects) to next-to-next-to-next-to leading order (NNNLO). (See \cite{Blanchetreview} for a recent review.) The conservative part of the dynamics has been tackled independently by several groups using different techniques, including the computation of the ADM Hamiltonian \cite{schafer3pn}, the equations of motion in harmonic coordinates \cite{blanchet3pn}, and the two-body Lagrangian \cite{eft3pn} using the Effective Field Theory (EFT) framework introduced in \cite{nrgr}. (The EFT constructed in \cite{nrgr} was coined NRGR due to similarities with EFTs in heavy quark physics.}) 
The analytic computation of the binary dynamics to 4PN order is underway, and progress has been reported during the conference by two independent groups, the NRGR \cite{nrgr4pn} and ADM \cite{adm4pn} teams, represented by {\it R. Sturani} and {\it P. Jaranowski} respectively. \\
 
All these derivations are obtained using a point-particle approximation for the constituents of the binary, which entails regularizing divergences that appear due to the uses of $\delta$-like sources.  The decoupling of internal structure in the dynamics of the binary is often assumed to hold to 5PN order for non-spinning objects. This is the so-called {\it Effacement Theorem}. The proof has been discussed in different approaches, and it is most clearly understood within the EFT formalism, where divergences of the point-particle approximation are tackled by dimensional regularization and renormalized by the introduction of higher derivatives (non-minimal) couplings in the worldline action of the body, which account for their finite sizes \cite{nrgr}. Using the standard power counting rules of the EFT one can show the first such term appears at ${\cal O}(v^{10})$ once incorporated in the dynamics of the binary with spinless bodies \cite{nrgr}. Moreover, for the case of black holes it has been argued that many of these new terms (electric-type), including the one at 5PN, have vanishing (renormalized) coefficients in four space-time dimensions, but in general do not vanish in higher dimensions \cite{smolkin}.\footnote{It is still possible these terms may be needed as pure {\it counter-terms} to regularize divergences.} This result was presented by {\it M. Smolkin}, and implies that finite size effects for black hole binaries must enter at higher orders (magnetic-type). For other type of objects, such as neutron stars, these terms ({\it aka} Love numbers) do not vanish and instead encode the corrections due to the short distance physics that enters in the equation of state, formally starting at 5PN order, although expected to be numerically enhanced \cite{Flan,nagar, poisson}. This means GW observations will probe the inner structure of neutron stars. Progress towards a complete description of these effects in the worldline approach of NRGR \cite{dis1,dis2}, and obtaining the (Wilson) coefficients of these new terms form the physics of neutron stars, was reported by {\it J.~Steinhoff}~\cite{love}.\footnote{One somewhat intriguing aspect of these computations is the fact that the (dissipative) imaginary part of the response function for the mass multipole moments of a non-rotating black hole (in four space-time dimensions) to a gravitational external perturbation does not vanish \cite{dis1}, unlike its real (conservative) part. This implies the traditional (unsubtracted) dispersion relation connecting real and imaginary parts of Green's functions is not valid. (Something similar occurs in describing dissipative effects in the EFT of fluids \cite{fluids} using the methods developed in \cite{dis1,dis2}, which may not be unrelated in light of the {\it Membrane Paradigm}.)}\\

The recent observations (e.g. \cite{spin}) which suggest compact objects in binary systems as well as supermassive black holes may be rapidly rotating, and the exciting possibility to test the most {\it twisted} properties of General Relativity, has motivated the study of spin effects in the GW waveforms. The parameter estimation from GW waveform including spin was discussed by {\it A. Nielsen} \cite{nielsen}, and other aspects of GW detection was presented by {\it A. Gupta} \cite{gupta}.

Allowing the bodies to rotate complicates matters significantly, and different approaches have been pursued to study spin effects in General Relativity; most notably the extensions of NRGR and the ADM canonical formalism to spinning bodies \cite{nrgrs, admspin}, as well as computations in harmonic gauge \cite{Blanchetreview}. The leading order effects in the dynamics linear in the spin at 1.5PN order were first obtained in the 70's \cite{barker}.
The NLO terms were (much) later computed in \cite{owen, faye} and re-derived in \cite{admso} and \cite{nrgrso, levi}. The NNLO equations of motion linear in spin have been computed independently in harmonic \cite{marsat} and the ADM formalism \cite{steinh}, the details of the former were reported by~{\it S.~Marsat}.

At quadratic order in the spin not only one encounters spin-spin interactions between the constituents of the binary, but also spin$^2$ terms that encode finite size effects, such as the intrinsic quadrupole moment of a spinning black hole. Unlike spinless bodies, there is no effacement of internal structure for spinning objects since spin$^2$ terms already appear at 2PN, namely at the same order as the leading spin-spin interaction, computed in \cite{barker} (see also \cite{wald}). Finite size effects due to spin can be readily incorporated in the EFT framework developed in \cite{nrgrs} where a worldline effective action approach for spinning bodies in General Relativity was constructed, including higher derivative terms encoding the extendedness of the compact objects. Using the power counting rules of NRGR it is simple to show that {\it one and only one} new term is required to 3PN order, whose Wilson coefficient can be determined by matching the metric of an isolated rotating compact object in the full theory and EFT sides. This highlights some of the simplifications behind the EFT approach, rather than working at the level of the field equations, where many (a priori independent) contributions to the stress-energy tensor of a spinning extended object can be shown to derive from the same (finite size) term in the effective action, a scalar under the symmetries. Using the (Feynman) rules derived in \cite{nrgrs} the NLO spin-spin and spin$^2$ gravitational potentials for spinning compact bodies were obtained in \cite{s1s2,nrgrss,nrgrs2}, as well as the equivalent ADM Hamiltonians computed in \cite{admss,adms2,adms2n}. Full agreement between these results has been reported \cite{agree}. To date the NNLO spin-spin potential/Hamiltonian has been computed in NRGR \cite{levi2} and ADM \cite{steinh2} frameworks.

The computation of the GW amplitude and phase require obtaining the energy loss in GW emission, which in turn is decomposed in multipole moments. (See \cite{Blanchetreview} and \cite{andi} for a derivation in the more traditional and EFT frameworks.) The radiative multipole moments necessary to account for effects linear in spin in the waveforms to NLO were obtained in \cite{faye2}, and \cite{eftrad} in NRGR, and to NNLO in \cite{bohe}. The details of the latter were reported by {\it A.~Bohe}. To date, only the NRGR formalism has succeeded in computing the radiative multipole moments necessary to include spin-spin and spin$^2$ terms to 3PN order \cite{eftrad}, as well as higher order spin dependent multipoles for the amplitude, up to 2.5PN order \cite{eftamp}. Progress towards obtaining the GW waveforms including all spin effects to 3PN order was the subject of {\it A. Ross} presentation. 

One important contribution to the GW emission is the so-called tail effects, or scattering of the emitted GW off the binaries background geometry. For spinning bodies the leading order tail contribution linear in spin enters at 3PN \cite{tailbuo1}, and the NLO contribution has been computed in \cite{tailbuo2} and also reported by 
{\it A. Marsat}. The tail effect introduces novel features, such as the renormalization of the multipole moments and subsequent renormalization group structure of NRGR \cite{andiwalter}, which allows to resum certain logarithmic UV (short distance) corrections. More generally, the renormalization group structure of the terms in the long-distance effective action allows one not only to resum UV logs, but also to identify logarithmic contribution to the conservative sector, such as the leading logarithmic term to the binding energy at 4PN for spinless bodies \cite{massnrgr}, derived in \cite{letiec} together with the NLO contribution at~5PN. \newpage

Recently (analytic and numerical) computations of the self-force have received significant attention in the PN community after some higher order PN corrections have been shown to derive from the former at leading order in the mass ratio \cite{letiec,det,letiec0,letiec2,firstlaw}. This was part of the presentation of {\it J. L. Friedman}, who reported on the status of self-force computations for EMRIs. Remarkably, a {\it conservative} 5.5PN term has been found in the expansion of the binding energy of the binary \cite{whit}, and higher order effects have already been computed \cite{damour} (see also \cite{foffa}). As discussed by {\it G. Faye} at the conference, this term derives from the `tail-of-tail' contribution to the stress-energy tensor \cite{tailfaye}. This  correspond to a higher order term in the analysis of \cite{massnrgr}. In general, conservative and dissipative terms appear at $n$PN order with $n$ even and odd respectively for spinless bodies. (Spin changes the parity properties of the equations.) This is related to the time-reversal properties of the different terms. The appearance of a 5.5PN term demonstrates the subtleties of the problem, which it is ultimately dissipative in nature. As discussed in \cite{massnrgr} an emitted GW can scatter back off the geometry of the binary which means, for long distance observables, there is a renormalization of the mass/energy of the system. This can therefore accommodate odd PN effects into the {\it conservative} quantities found in \cite{letiec, det,letiec0,letiec2,firstlaw,whit}, defined in this way.

In a priori completely different regime, the self-force computations for EMRIs is expected to provide templates for space-based GW observatories operating at much lower frequencies than LIGO/Virgo. Many of the subtleties of computing the waveforms for EMRIs and other aspects of black hole perturbation theory were discussed during the conference, with talks by {\it C. Merlin} \cite{merlin}, {\it A. Heffernan} \cite{anna} and {\it B. Nolan} \cite{nolan}, on the regularization issues of the self-force, and {\it T. Hinderer} \cite{tanja} and {\it R. Cole} on the importance of resonances. 

Analytic computations of the gravitational self-force are difficult, and up to date only second order effects in the mass ratio, i.e. ${\cal O}(q^2)$ with $q\equiv m/M \ll 1$, are (formally) known \cite{pound,gralla}. In principle to compare with PN effects the self-force computations referred above should be applied to EMRIs in weak field configurations in non-relativistic motion.
In order to relate self-force with comparable mass PN or numerical results, the following replacement is often performed $q \to \nu \equiv \tfrac{m_1m_2}{(m_1+m_2)^2}$, which is accurate to leading order in the mass ratio for EMRIs. Many of these comparisons, most notably with numerical results, are obtained using the self-force to ${\cal O}(\nu)$. In his talk {\it A. Le Tiec} showed that a remarkable agreement is found already at leading order even for comparable mass inspirals, when $\nu \simeq 1/4$ \cite{peri0,bind,peri}. 

This may suggest another peculiar feature of gravitational dynamics, for instance for the expansion of the binding energy \cite{bind} (and other observables such as the total angular momentum, periastron advance, etc.) in terms of the relative velocity and symmetric mass ratio, where effects naively down by a factor of $\nu \simeq 1/4$ may be further suppressed. It is known ${\cal O}(\nu^2)$ terms start at 2PN order, and therefore kick in for relativistic motion. However, the agreement remains surprisingly faithful even in the strong gravitational regimes, for $v \simeq 1/2$ \cite{bind}. This may be related to the `Unreasonable Effectiveness of Post-Newtonian Theory' \cite{Will}, and might entail cancelations, perhaps of the same type encountered in other computations in gravity~\cite{bern}, see also \cite{duff}. (It may also be related to the vanishing of the electric-type finite size effects for binary black holes \cite{smolkin,nagar,poisson}.)\\

Obtaining higher order self-force effects is therefore of great relevance. One example in which simplifications arise is the ultra-relativistic limit of the self-force problem. As discussed by {\it C.~Galley} at the conference one can re-organize the standard perturbative expansion in the mass ratio $q$ within the EFT formalism applied to EMRIs \cite{chad} in powers of $\lambda= N \epsilon$, where $N = 1/\gamma^2$ and $\epsilon =\gamma q$ with $\gamma$ the boost factor \cite{largeN}. Using the power counting rules developed in \cite{largeN} one can show the large $N$ limit reduces the number of terms significantly, similarly to what occurs in gauge theories \cite{tony}, and higher order effects can be readily obtained at ${\cal O}(1/N)$. For instance, in \cite{largeN} the computation of the self-force was carried up to fourth order in $\lambda$ at leading order in $1/N$, while the regularization of the divergences of the point-particle approximation become trivial in dimensional regularization\footnote{Scaleless integrals are set to zero in dimensional regularization.} since higher order terms in the effective action are suppressed in the large $N$ limit. The computation of the self-force in the EFT approach entails a subtle use of the {\it in-in} formalism in a classical setting, as discussed by {\it C. Galley} \cite{adam,chad2}. This was also the topic of {\it B. Kol}~'s talk \cite{kol}.

Understanding different corners of the perturbative expansion is essential to unravel the underlying features of gravitational dynamics. An attempt to produce analytical waveforms that can be applied to different regimes, used to scan the different parameters of the problem and perhaps shed some light on these matters, including the strong gravitational realm, is the so-called Effective One Body (EOB) approach \cite{EOB}. Different aspect of the EOB paradigm to describe the inspiral, merger and ringdown phases were discussed during the conference, most notably for spinning binary systems, with talks by {\it Y. Pan} and {\it A. Taracchini} \cite{pan,tara}. EOB waveforms are calibrated with numerical counterparts. Numerical techniques have matured into a very successful area of research (see the proceedings for session B2). As part of the study of binary systems, although without yet a full control of all the cycles, numerical templates are useful not only to describe the merger but also to match models for the waveforms. Hence the meeting also incorporated a joint session B2/B4 which had (among others described above) reports on the status of numerical methods in General Relativity and hybrid approaches. In particular the talks by {\it H. Pfeiffer} \cite{harald} and {\it S.  Husa} \cite{husa} on numerical simulations for binary black holes,  {\it G. Lovelace} \cite{lovelace} on simulations for compact binaries with nearly extremal black-hole spins, and {\it S. Kahn}  on the structure of ringdown modes \cite{Kama}.

\subsection{Modified Theories of Gravity and Fundamental Issues}

One hundred years of Einstein's gravity have introduced a potentially dangerous bias towards a theory which may
breakdown somewhere between the six-orders of magnitude difference in gravitational potential at the surface of the Sun and at the surface of a neutron star or black hole. Thus, a considerable amount of intellectual effort is being channelled into understanding the consequences of modified theories of gravity and how they may affect gravitational dynamics. One of the most popular models to modify the field equations includes scalar-tensor theories, which give rise to novel effects in the presence of matter~\cite{Damour:1992we,Cardoso:2011xi,Cardoso:2013opa}, but can also affect vacuum spacetimes. Breakdown of no-hair theorems and scalar-emission by black hole binaries in these theories was discussed by {\it L. Gualtieri} \cite{gual}. Other examples of quadratic theories are Gauss-Bonnet~\cite{Pani:2009wy} and Chern-Simons gravity~\cite{JP, Alexander:2009tp}. 
Limits on the latter coming from GW and pulsar observation probes were discussed by {\it K. Yagi} and {\it L.~Stein} \cite{cs1,cs2}. (For~\,other type of --more fundamental-- constraints see \cite{marc}.) On the other hand, {\it P. Pani} discussed perturbations of slowly rotating black holes \cite{Pani} as well as extensions of General Relativity including minimally coupled massive fields, and how these well-motivated theories can yield interesting smoking-gun effects in strong-field gravity. A particularly interesting consequence is the resulting competitive bounds on the photon mass from observations of supermassive black holes~\cite{Pani:2012vp}.\\

The sessions were completed with the application of perturbation theory to other fundamental issues in gravity. 
It was recently conjectured that the event horizon of black holes (and cosmic censorship) could be destroyed by throwing point particles at charged or spinning black holes~\cite{Hubeny:1998ga,Jacobson:2009kt}. {\it J. Rocha} discussed the extension of these results to higher dimensional and asymptotically anti-de Sitter spacetimes~\cite{rocha,cardoso}.
Such processes neglect conservative self-force effects, which have been conjectured to prevent destruction of the horizon
and therefore to preserve cosmic censorship~\cite{Barausse:2010ka}.
{\it M. Colleoni} described on-going efforts to analyse rigorously self-force effects in such challenging spacetimes and on the possibility that self-force prevents overspinning a Kerr black hole. {\it J. Camps} discussed an important perturbation-theory result concerning the Gregory-Laflamme instability \cite{camps}; {\it N. Warburton} discussed iso-frequency pairing of geodesic orbits for Kerr black holes \cite{niels}  and {\it O. Moreschi} talked about properties of a `Particle Model' to describe compact objects in the null gauge~\cite{more}. Finally, the computations reported in \cite{largeN} in the ultra-relativistic limit may also be relevant to study the high-energy behavior of scattering amplitudes, and the `S-matrix' for gravity \cite{smatrix}. Features of trans-Planckian gravitational scattering were also discussed by {\it D. Gal'tsov}~\cite{galtsov}.

\begin{center}
{\bf Acknowledgments}
\end{center}
We thank all the participants of B4 and B2/B4 at GR20 \& Amaldi 10, as well as the organizers of such a wonderful conference, especially B. Iyer for inviting us to chair these sessions. V.~C. acknowledges financial support provided under the European Union's FP7 ERC Starting Grant ``The dynamics of black holes: testing the limits of Einstein's theory'' grant agreement no. DyBHo--256667. R.A.P. acknowledges support by the NSF grant AST-0807444 and the  DOE grant DE-FG02-90ER40542 at the IAS, and by the German Science Foundation within the Collaborative Research Center 676 `Particles, Strings and the Early Universe,' at DESY.

\end{document}